\documentclass{pasj00}
\usepackage{graphicx}
\usepackage{epsfig}
\usepackage{lscape}
\draft
\begin{document}

\SetRunningHead{S. S. Weng et al.}{Running Head}
\Received{}
\Accepted{}

\title{Spectral Analyses of the Nearest Persistent Ultraluminous X-Ray Source M33 X-8}

 \author{%
   Shan-Shan \textsc{Weng}\altaffilmark{1},
   Jun-Xian \textsc{Wang}\altaffilmark{2},
   Wei-Min \textsc{Gu}\altaffilmark{1}
  and
   Ju-Fu \textsc{Lu}\altaffilmark{1}
}
 \altaffiltext{1}{Department of Physics and Institute of Theoretical Physics
and Astrophysics, \\ Xiamen University, Xiamen, Fujian 361005,
P.R.China}
 \email{guwm@xmu.edu.cn}
 \altaffiltext{2}{Center for Astrophysics, University of Science and
Technology of China, \\ Hefei, Anhui 230026, P.R.China}

\KeyWords {accretion, accretion disks --- black hole physics ---
X-rays: binaries --- X-rays: stars --- X-rays: individual (M33
X-8)}

\maketitle

\begin{abstract}

We provide a detailed analysis of 12 XMM observations of the
nearest persistent extragalactic ultraluminous X-ray source (ULX),
M33 X-8. No significant spectral evolution is detected between the
observations, therefore we combine the individual observations to
increase the signal-to-noise ratio for spectral fitting. The
combined spectra are best fitted by the self-consistent $p$-free
disk plus power-law component model with $p$ =
$0.571_{-0.030}^{+0.032}$, $kT_{\rm in}$ = $1.38_{-0.08}^{+0.09}$
keV, and the flux ratio of the $p$-free disk component to the
power-law component being 0.63:0.37 in the 0.3 -- 10 keV band. The
fitting indicates that the black hole in M33 X-8 is of $\sim 10
M_{\rm \odot}$ and accretes at a super-Eddington rate ($\sim$ 1.5
$L_{\rm Edd}$), and the phase of the accretion disk is close to a
slim disk ($p$ = 0.5). We report, for the first time, that an
extra power-law component is required in addition to the $p$-free
disk model for ULXs. In super-Eddington cases, the power-law
component may possibly result from the optically thin inner region
of the disk or a comptonized corona similar to that of a standard
thin disk.

\end{abstract}

\section{Introduction}
Ultraluminous X-ray sources (ULXs) are point-like, non-nuclear
X-ray sources with isotropic luminosities of about
$10^{39}-10^{41}$ ergs s $^{-1}$ in nearby galaxies \citep{fab89}.
These objects are interesting since their luminosities are
intermediate between the luminosities of Seyfert galaxies ($L_{\rm
X}$ $\sim $ $10^{42}-10^{44}$ ergs s $^{-1}$) and those of black
hole X-ray binaries (BH XRBs; typically $L_{\rm X}$ $\lesssim$
$10^{38}$ ergs s $^{-1}$). If X-rays are emitted isotropically
below the Eddington limit, the intermediate luminosity would
indicate that ULXs harbor intermediate-mass black holes (IMBHs;
$M_{\rm BH}$ $\sim $ 20--10$^3$ $M_{\rm \odot}$). However, IMBHs
are not required if the emission is anisotropic or
relativistically beamed \citep{king01, king09, kor02}, or the
accretion is super-Eddingtonian by a factor of a few
\citep{beg02,po07}.

X-ray spectral fitting is one of the best methods known to weigh
the black hole in ULXs while the dynamical method is unavailable.
The sum of a multicolor disk (MCD; Shakura \& Sunyaev 1973) and a
power-law (PL) model is widely used to describe the X-ray spectra
of black hole binaries. This canonical MCD$+$PL model reflects the
expectation of the thermal emission from a standard thin disk
around a Schwarzschild black hole along with the hard emission
from the inverse Compton scattering of disk photons. Fitting ULX
spectra with the same canonical model often shows that the PL
component dominates the 0.3-10.0 keV spectrum and the disk
component characterizes the feature of the soft spectrum
significantly below 1 keV \citep{mil03,gs06}. The disk temperature
is inversely related to the black hole mass ($T_{\rm eff} \propto
M^{-1/4}$; e.g., Makishima et al. 2000) in a MCD model, i.e., a
heavier black hole tends to accrete with a cooler disk. The
obtained disk temperature around 0.15 - 0.2 keV in many ULX
spectra was used to suggest the existence of IMBHs with masses of
$\sim 10^{3} M_{\rm \odot}$ \citep{kar03,la09}.

Meanwhile, hotter disks  with temperature $kT_{\rm in}$ $\sim$
1-2.5 keV were also reported in literature. In these cases, the
disk component dominates the X-ray emission \citep{sto06}, in a
way similar to Galactic BH candidates. The hot MCD model (HD
model) with fitted temperature $kT_{\rm in}$ $\sim$ 1-2.5 keV
would imply super-Eddington luminosities and stellar-mass black
holes in the ULXs. However, the MCD model is based on the standard
thin disk model and is valid only for luminosities well below
$L_{\rm Edd}$ ($\lesssim$ 0.1 $L_{\rm Edd}$). It is known that the
radial temperature of a standard thin disk follows a power-law
form as $kT \propto r^{-0.75}$. Such a form is derived from the
assumption of energy balance between the viscous heating and the
radiative cooling. For sufficiently high accretion rates, the
radiative cooling itself can not balance the viscous heating and
the advective cooling becomes important or even dominant. In this
case, the power-law form $kT \propto r^{-0.75}$ is invalid. In
fact, there are observed deviations from the standard disk
spectrum in BH XRBs as they approach their Eddington limit
\citep{kub04}.  In the slim disk model, that is applicable to
super-Eddington accretion \citep{ab88}, the radial temperature
follows $kT \propto r^{-0.5}$. This demonstrates that the HD model
is not self-consistent for ULXs \citep{gs06}.

The so-called $p$-free model also obeys a power-law form as
$T_{\rm eff}\propto r^{-p}$, but the temperature gradient $p$ is
allowed to vary from 0.5 to 0.75 \citep{min94}. The $p$-free model
is an extended MCD model and is a combination of two  types  of
optically  thick  disk models, namely the  standard thin  disk and
the  slim disk. \citet{kiki06} successfully applied the $p$-free
model to the spectra of four ULXs, which were previously reported
to contain cool disks and thus IMBHs. The fitting yields $p$
values of $\sim$ 0.5, consistent with the slim disk model,
suggesting that the black holes in these four ULXs have stellar
masses and accrete at super-Eddington rates instead. Note that in
their fitting, no extra PL component was considered.

The ULX source M33 X-8 with an X-ray luminosity of $\sim$
$10^{39}$ ergs s$^{-1}$ was discovered by \citet{long81} with the
Einstein satellite. Although the position of M33 X-8 coincides
with the optical center of the galaxy  \citep{la03}, the
hypothesis of an active galactic nucleus (AGN) is inconsistent
with the estimated upper limit of 1500 $M_{\rm\odot}$ on the
central black hole mass  in M33 \citep{geb01}; moreover, no AGN
activity has been found in other bands for this source. The source
is of particular interest for many reasons. First, its X-ray
spectra apparently prefer the HD model, but the model is unlikely
to be appropriate at this luminosity. Second, it is the nearest
persistent extragalactic ULX \citep{fo04}. Third, it had up to 12
observation data available in the XMM-Newton Public Archive. In
this paper, we present a detailed analysis of the 12 XMM exposures
on M33 X-8, especially the combined spectra, which have much
higher signal-to-noise (S/N) ratio and enables us to provide
further constraints on the nature of this source.

\section{Data Reduction}
In Table 1 we list all the 12 XMM observations of M33 X-8, which
were obtained from August 2000 to July 2003. Hereafter we refer
them as Obs \#1 through Obs \#12 for convenience. The data were
reduced with the XMM-SAS software version 7.1.0. To exclude
intervals with background flares, we created light curves for
photons above 10 keV, and used a count rate cut-off criterion to
filter the light curves. The exact value of the cut-off was
allowed to vary from field-to-field, to provide the best
compromise in each case between excluding high background periods
and facilitating the longest available exposure \citep{sto06}. We
selected the data from good time intervals, by setting FLAG = 0
and PATTERN $\leq$ 4 for PN data, and PATTERN $\leq$ 12 for MOS
data. The source spectra were extracted from circles with radius
of 35\arcsec\ and centered at the nominal position of M33 X-8 (RA
= \timeform{01h33m50s.89}, Dec = +\timeform{30D39'37".2}, J2000),
while the background spectra were extracted from the same CCD
chips as the source and at a similar distance from the readout
node. The high spatial resolution Chandra image has confirmed that
there is no obvious contamination from neighbouring point sources
within several arcminutes \citep{dub04}. For Obs \#1 that the MOS
camera was operated in small window mode, we used the background
in the closest chip. With the SAS task epatplot, we found that
three observations (Obs \#2, \#4, and \#8) were affected by
pile-up. Only PN data for these three observations were used and
the spectra were extracted in annulus regions with radius
15\arcsec\ and 40\arcsec\ to circumvent the effects of pile-up
\citep{xmm}. In these cases, ARF files were calculated to correct
the missing part of PSF, thus the correct flux level could still
be measured through spectra fitting. In several cases, PN or MOS
data were unavailable because the source was not covered by the
detector, or due to CCD gaps. When available, PN and MOS data from
each individual observation were fitted together for spectral
analyses. The 0.3 -- 10.0 keV band spectra were fitted with the
HEAsoft X-ray spectral fitting package XSPEC 12.3.1.  All spectra
were rebinned to have at least 20 counts per bin to enable the use
of $\chi^2$ statistics.

\section{Spectral fitting}
\subsection{Individual observations}
We first fit the 12 individual spectra with an absorbed MCD$+$PL
model (diskbb$+$po in XSPEC) and a $p$-free disk model (diskpbb),
respectively. We get similarly good fits for both models (Table
1). During the fitting, the absorption column density is allowed
to vary \citep{feng06}. Due to the limited number of photons, it
is difficult to judge whether the MCD$+$PL model or the slim disk
model is preferred. However, by examining the fitting residuals of
the $p$-free model, in three of the observations with the highest
number of data bins (Obs \#1, \#5, and \#12), we find a weak hard
tail above 7 keV, which we will further investigate for the
composite spectra in \S3.2.

M33 X-8 was reported as a persistent source by \citet{fo04}, and a
small flux variation was detected previously for a modulation of
$\sim$ 20\% with a period of 106 days \citep{dub97}. Consistently,
we obtain that the amplitude of the flux variation between the
observations is at $<20\%$ level, except for Obs \#4, and \#8,
which are about 1.5 -- 1.7 times brighter than the average level
of the other 10 observations. In Figure 1 we plot the X-ray light
curve  for the 12 observations of M33 X-8. The 0.3 -- 10.0 keV
luminosities are taken from the best-fitted MCD$+$PL model. While
different models yield slightly different luminosities, the
general pattern of the light curve will not be changed. The mean
luminosity (with Obs \#4 and \#8 excluded) along with the ratio of
the luminosities to the mean value are also shown. From the figure
 it is seen that  M33 X-8 remains persistent during most of the
observations, and we detected no significant spectral evolution
through spectral fitting (see Table 1). No significant rapid
variation was detected either within individual observations.

\subsection{Combined spectra}
Inspired by the fact of no significant spectral evolution between
the individual observations, we combined all the observations
except for Obs \#4 and \#8, and obtained one co-added PN spectrum,
one co-added MOS1 and one MOS2 spectrum, respectively, in order to
increase the S/N ratio of the spectra. The co-adding was performed
with the FTOOLS ``addspec" which adds pulse-height amplitude (PHA)
spectra and background PHA files; detector redistribution and
ancillary response were also combined with source net photon
counts detected in each observation as co-adding weight. We fit
the co-added PN, MOS1 and MOS2  spectra simultaneously. We
excluded Obs \#4 and \#8 because they are at significantly higher
flux level, however, we have checked that including them does not
change our results in this paper.

We apply the same models mentioned in \S3.1 to the combined high
S/N spectra to test whether these models can still provide
adequate fits. Neither the HD model nor the simple $p$-free model
can give good fits below $\sim$ 1 keV (see Figs. 2a and 2b).
\citet{la03} found that a thermal plasma component is required to
represent the extended emission around the point source (also see
Schulman \& Bregman 1995). Accordingly, we add a Raymond-Smith
component to the above models, then the $\chi^{2}$ is reduced by
more than 50 with 3 additional free parameters, yielding an F-test
probability lower than $10^{-9}$.  After including the
Raymond-Smith component, both models (HD and $p$-free) yield
similar $\chi^2$, and the fitting results are also listed in Table
1. Since the HD model ($kT_{\rm in}$ $\sim$ 1.16 keV) is known to
be inconsistent for ULXs, we focus on the $p$-free model below.

Upon investigating the fitting carefully (see Figs. 2b and 2c), we
see that the $p$-free disk model (plus the Raymond-Smith component
or not) can not provide an acceptable fit to the band above 7 keV.
The hard tail above 7 keV shown in the plot suggests the existence
of an additional hard component. Following  previous works on BH
XRBs (e.g., Kubota \& Makishima 2004), we add  to the $p$-free
disk model an extra power-law component (Fig. 2$d$), and the fit
is significantly improved with a confidence level above 99.99999\%
based on the F-Test.  Noting that the F-test was questioned in
testing the significance of an additional spectral component
\citep{Pro02}, we performed Monte-Carlo simulations to demonstrate
the significance of the power-law component.  In terms of the
best-fitted model without a power-law component, we made 1000
artificial spectra, and run spectral fitting by adding an extra
power-law component. In the 1000 simulations, the extra power-law
component can only improve the fitting with $\Delta\chi^2$ $<$ 16,
far below the actual $\Delta\chi^2$ = 38.65 in the real spectra,
which means that the confidence level is far above 99.9\%.
Therefore, the extra power-law component is statistically solid,
and the same statement applies to the Raymond-Smith component
stated in the above paragraph.

Furthermore, with the following reasons we can rule out the
possibility that the hard tail in the spectra is due to improper
background subtraction. First of all, the hard tail is not only
visible in the co-added spectra, it is also obvious while we fit
all the individual PN, MOS1 and MOS2 spectra together. This
implies that the hard tail is not due to the possible improper
background subtraction during the spectrum co-adding.  Secondly,
the hard tail is also visible in three of the individual
observations with the highest number of data bins.  Spectral
fitting  to individual observations  also rules out  the
possibility  that the  hard tail  is dominated  by one single
observation.

\section{Discussion}
Various models have been proposed to explain the spectra of ULXs
in literature. However, due to limited photon counts, in many
cases it is not possible to distinguish these models by spectral
fitting only. Extra constraints on these models have been given by
analyzing the spectral evolution, as reported for several ULXs,
e.g., IC 342 \citep{kubo01}, NGC 1365 X-1 \citep{sor07}, and
Holmberg IX X-1 \citep{la01}. After investigating the spectral
evolution of NGC 1313 X-2, \citet{feng07} found that the variation
of the accretion disk component deviated significantly from $L
\propto T^{4}$ relation if fitted with MCD$+$PL model, while
roughly consistent with $L \propto T^{4}$ if fitted with the
$p$-free model. They suggested that this source supports the slim
disk model and is against the MCD model. However, flux variations
by more than one order of magnitude are rare in ULXs, thus such a
technique can not be applied to most ULXs, such as M33 X-8 we
studied here, which shows only small amplitude variations.

In this paper we present detailed X-ray spectral fitting to 12 XMM
exposures on the ULX source M33 X-8. We found no significant rapid
variations within individual observations. The X-ray flux remains
persistent during 10 of these 12 observations with the flux
variation amplitude $<$ 20\%, and no significant spectral
evolution was detected between observations. Then the data of
these 10 observations were combined to derive composite spectra
with much higher S/N ratio for spectrum fitting. We find that the
MCD$+$PL model and the $p$-free disk model provide comparable fits
to both individual and composite spectra.

However, a significant hard tail above 7 keV is detected in the
residual spectra of the $p$-free disk model. We added a power-law
component to the $p$-free disk model to represent this hard tail,
and find that the $p$-free disk $+$ PL model provides the
statistically best fit comparing to other models. The probable
reason for the power-law component in this model is that the inner
disk becomes optically thin in super-Eddington cases (Artemova et
al. 2006), or there is a comptonized corona \citep{gl09} similar
to that of a standard thin accretion disk. The fluxes of the
$p$-free disk component and the power-law component  in 0.3--10.0
keV band are 1.43$\times10^{-11}$ ergs cm$^{-2}$ s$^{-1}$ and
0.83$\times10^{-11}$ ergs cm$^{-2}$ s$^{-1}$, respectively, with a
ratio of  0.63:0.37; this indicates that the $p$-free disk
component dominates over the power-law component.

The X-ray luminosity of M33 X-8 is $1.7\times10^{39}$ ergs
s$^{-1}$ for the given distance of 0.7 Mpc \citep{ho97}. The
luminosity and the best-fitted inner disk temperature $kT_{\rm in}
= 1.38$ keV indicate that the mass of the black hole in M33 X-8 is
$\sim 10 M_{\rm \odot}$ (see Fig. 1 of Watarai et al. 2001), in
good agreement with previous works \citep{mak2000,fo04}. Because
the X-ray emission is dominated by the disk component, the mass
derived here should be reliable \citep{kiki06}. We note that the
luminosity exceeds marginally the Eddington luminosity of a $\sim
10 M_{\rm \odot}$ black hole. In this case, the X-ray emission
from the outer region of a standard thin disk could not be
neglected, and the moderate value of $p$ = 0.571 is well
consistent with the theoretical calculation \citep{wat00}. In
Galactic black hole X-ray binaries, the very high state (or steep
power-law state) is defined as a state in which the luminosity is
exceedingly high ($L_{\rm X}$  $>$ 0.2 $L_{\rm Edd}$ ) and
the X-ray spectrum displays substantial nonthermal radiation,
which may constitute 40-90 \% of the total flux, with a photon
index larger than 2.4 \citep{mc06}. The state of M33 X-8 is more
like the thermal-dominant state of X-ray binaries in the Galaxy,
but with much higher luminosity, occupying a new ultraluminous
accretion state (Gladstone et al. 2009).

\section*{ACKNOWLEDGMENTS}
We thank the referee for constructive comments, and Lu-Lu Fan and
Hua Feng for valuable discussions.  This work was supported by the
National Basic Research Program of China under grant 2009CB824800,
the National Natural Science Foundation of China under grants
10778711 and 10833002, and Program for New Century Excellent
Talents in University under grant 06-0559.

\clearpage
\begin{figure}
\centerline{\psfig{figure=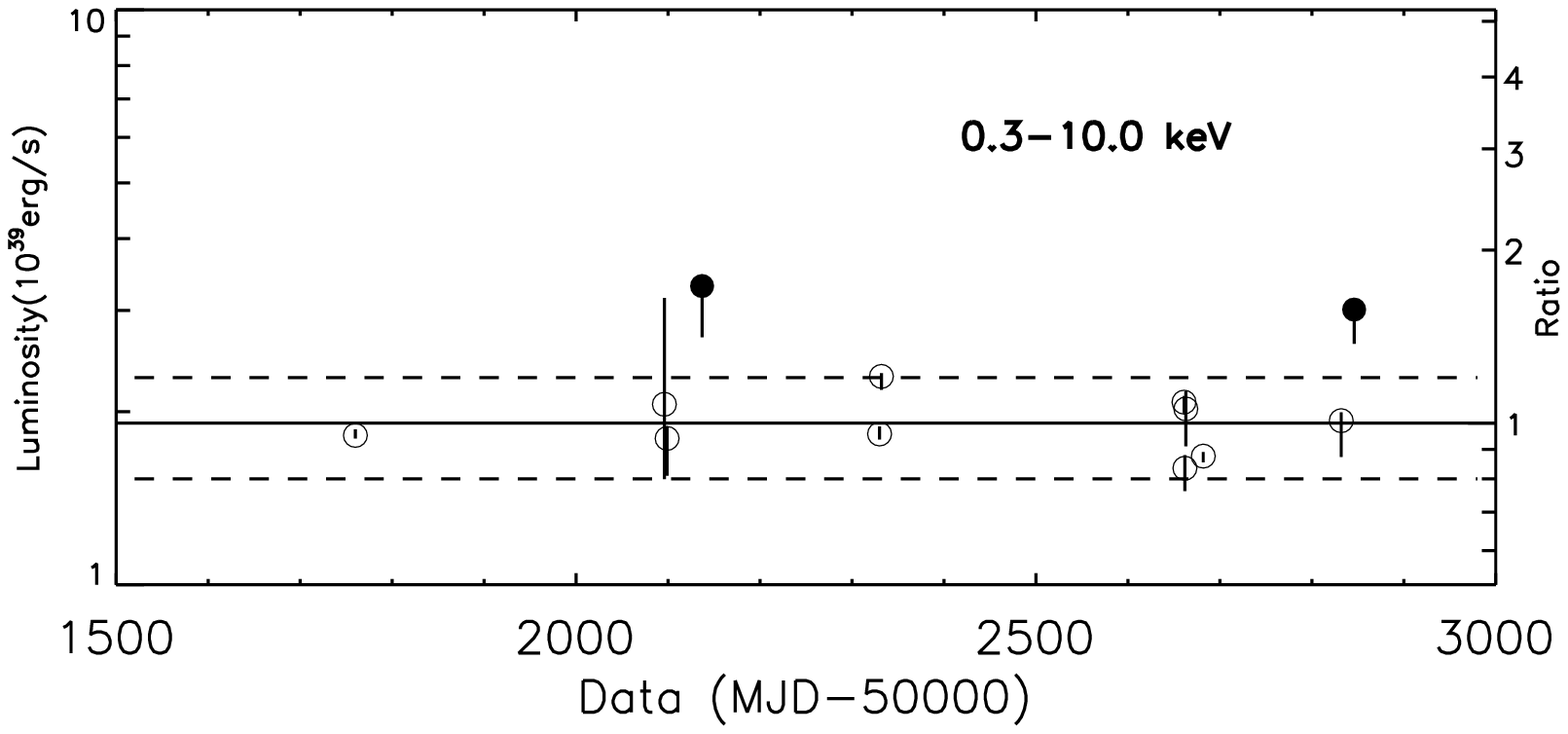,height=10.0cm,angle=0}}
\caption{The 0.3 -- 10.0 keV lightcurve of M33 X-8. Obs \#4 and
\#8, which show significantly higher luminosity level are marked
as solid circles. The solid line marks the mean luminosity of the
rest 10 observations. 120\% and 80\% of the mean luminosity are
marked by the dashed lines to illustrate the amplitude of the
variation. \label{fig:1}}
\end{figure}

\clearpage
\begin{figure}
\centerline{\psfig{figure=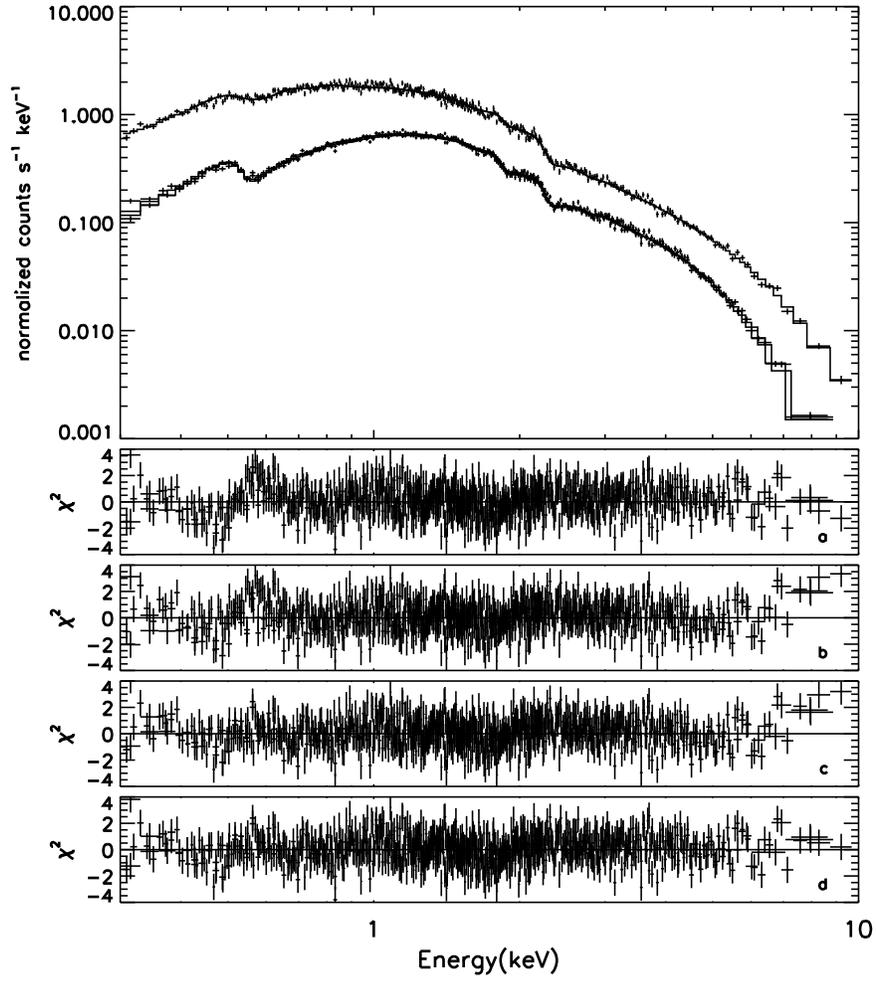,width=14.0cm,angle=0}}
\caption{Fitting to the combined spectra of ULX M33 X-8. Panels
($a$) and ($b$) show the fitting residuals for the HD model and
the $p$-free model, respectively. After adding a Raymond-Smith
component to the models, panels ($c$) and ($d$) show the residuals
for the $p$-free model and the $p$-free $+$ power-law model,
respectively. The combined spectra and the best fitted model (the
$p$-free $+$ power-law model) are plotted in the upper panel.
\label{fig:2}}
\end{figure}

\begin{center}
\begin{table*}
\centerline{\bf Table 1. BEST-FIT SPECTRAL PARAMETERS OF M33 X-8}
\tiny
\begin{tabular}{p{4mm}llllp{10mm}p{10mm}p{10mm}lcl}\hline
Obs & XMM ObsId & Instruments & Exposure & $N_{\rm H}$ & $kT$ &
$kT_{\rm
in}$ &  p  & $\Gamma$ & $f_{\rm X}$ & $\chi^2/$dof \\
 & &  & ksec & $10^{21}$ cm$^{-2}$ & keV & keV &  &
 & $10^{-11}$ ergs cm$^{-2}$ s$^{-1}$ &
\\ \hline
  & & & & &    phabs*(diskbb$+$po)     & & & &  \\
\hline

1 & 0102640101 & PN/M1/M2 & 7.4/10.5/10.5 &
$0.176_{-0.018}^{+0.017}$& & $1.08_{-0.04}^{+0.05}$  &   & $2.32_{-0.11}^{+0.11}$  & 2.49 & 1252.95/1164 \\

2 & 0102640601 & PN & 3.4  & $0.209_{-0.010}^{+0.014}$ &
 & $1.43_{-1.01}^{+1.77}$  &  & $2.42_{-0.67}^{+1.23}$  & 2.72 & 89.03/83 \\

3 & 0102641001 & PN/M1/M2 & 1.7/9.4/9.5 &
$0.184_{-0.024}^{+0.025}$ & & $1.28_{-0.29}^{+0.32}$ &   & $2.13_{-0.15}^{+0.22}$  & 2.43 & 570.91/577 \\

4 & 0102642001 & PN & 8.0 & $0.220_{-0.064}^{+0.115}$ &
  & $1.24_{-0.35}^{+0.20}$ &   & $2.68_{-0.52}^{+1.09}$ & 4.37 & 205.62/198 \\

5 & 0102642101 & PN/M1/M2 & 9.0/12.1/12.1 &
$0.175_{-0.022}^{+0.019}$ &  & $1.11_{-0.07}^{+0.07}$ &   & $2.27_{-0.15}^{+0.14}$ & 2.45 & 1073.01/1024 \\

6 & 0102642301 & M1/M2 & 12.1/12.1 & $0.246_{-0.035}^{+0.046}$ &
  & $1.20_{-0.12}^{+0.10}$ &   & $2.58_{-0.26}^{+0.41}$ & 3.05 & 424.32/466 \\

7 & 0141980101 & M1/M2 & 8.5/9.7 & $0.169_{-0.037}^{+0.048}$ &
& $1.02_{-0.11}^{+0.17}$ &   & $2.18_{-0.26}^{+0.43}$ & 2.26 & 406.91/381 \\

8 & 0141980301 & PN & 10.0 & $0.284_{-0.076}^{+0.095}$ &
& $1.17_{-0.09}^{+0.10}$ &   & $3.37_{-0.67}^{+0.78}$ & 3.97 & 234.86/240 \\

9 & 0141980401 & M1/M2 & 8.3/8.6 & $0.181_{-0.060}^{+0.040}$ &
& $1.03_{-0.17}^{+0.22}$ &   & $2.04_{-0.41}^{+0.29}$ & 2.81 & 417.66/374 \\

10 & 0141980501 & PN/M1/M2 & 1.5/7.9/8.1 &
$0.212_{-0.027}^{+0.029}$ & & $1.35_{-0.13}^{+0.11}$ &   & $2.40_{-0.19}^{+0.24}$ & 2.28 & 757.64/755 \\

11 & 0141980601 & PN & 10.5 & $0.172_{-0.045}^{+0.036}$ &
& $1.14_{-0.14}^{+0.14}$ &   & $2.22_{-0.30}^{+0.29}$ & 2.10 & 430.21/455 \\

12 & 0141980801 & PN/M1/M2 & 7.6/10.0/10.0 &
$0.192_{-0.011}^{+0.010}$ &  & $1.16_{-0.14}^{+0.17}$ &   & $2.22_{-0.05}^{+0.06}$ & 2.01 & 1297.30/1162 \\

 & Combined & PN/M1/M2 & 41.0/79.0/80.8 & $0.184_{-0.007}^{+0.006}$ &
  & $1.14_{-0.02}^{+0.03}$ &   & $2.27_{-0.04}^{+0.04}$  & 2.39 & 1627.55/1438  \\

\hline
 & & & & &    phabs*diskpbb    & & & &  \\
\hline

1 & 0102640101 & PN/M1/M2 & 7.4/10.5/10.5 &
$0.148_{-0.008}^{+0.010}$& & $1.43_{-0.04}^{+0.05}$ & $0.563_{-0.011}^{+0.009}$ &   & 2.26 & 1240.79/1166 \\

2 & 0102640601 & PN & 3.4 & $0.169_{-0.041}^{+0.022}$ &   &
$2.06_{-0.40}^{+0.91}$ & $0.505_{-0.005}^{+0.038}$ &   & 2.33 & 88.65/84 \\

3 & 0102641001 & PN/M1/M2 & 1.7/9.4/9.5 &
$0.166_{-0.013}^{+0.012}$ &  & $2.26_{-0.23}^{+0.28}$ & $0.522_{-0.009}^{+0.011}$ &   & 2.27 & 565.81/579 \\

4 & 0102642001 & PN & 8.0 & $0.162_{-0.025}^{+0.029}$ &
& $1.63_{-0.19}^{+0.42}$ & $0.519_{-0.019}^{+0.025}$ &  & 3.47  & 208.85/199 \\

5 & 0102642101 & PN/M1/M2 & 9.0/12.1/12.1 &
$0.148_{-0.011}^{+0.011}$&  & $1.50_{-0.06}^{+0.07}$ & $0.561_{-0.012}^{+0.013}$ &   & 2.23 & 1087.61/1026 \\

6 & 0102642301 & M1/M2 & 12.1/12.1 & $0.190_{-0.017}^{+0.016}$ &
  & $1.55_{-0.10}^{+0.11}$ & $0.537_{-0.014}^{+0.017}$ &   & 2.49 & 426.00/467 \\

7 & 0141980101 & M1/M2 & 8.5/9.7 & $0.173_{-0.022}^{+0.021}$ &
& $1.54_{-0.13}^{+0.16}$ & $0.535_{-0.018}^{+0.022}$  &   & 2.25 & 406.17/382 \\

8 & 0141980301 & PN & 10.0 & $0.156_{-0.030}^{+0.027}$ &
& $1.40_{-0.15}^{+0.20}$ & $0.531_{-0.026}^{+0.033}$ &   & 2.20 & 238.70/241 \\

9 & 0141980401 & M1/M2 & 8.3/8.6 & $0.180_{-0.024}^{+0.023}$ &
& $1.70_{-0.17}^{+0.24}$ & $0.543_{-0.020}^{+0.024}$ &   & 2.70  & 422.10/375 \\

10 & 0141980501 & PN/M1/M2 & 1.5/7.9/8.1 &
$0.162_{-0.013}^{+0.015}$ &  & $1.76_{-0.10}^{+0.13}$ & $0.549_{-0.013}^{+0.013}$ &   & 1.95 & 748.04/757 \\

11 & 0141980601 & PN & 10.5 & $0.147_{-0.018}^{+0.023}$ &
& $1.56_{-0.11}^{+0.18}$ & $0.563_{-0.024}^{+0.022}$ &   & 1.93 & 428.83/456 \\

12 & 0141980801 & PN/M1/M2 & 7.6/10.0/10.0 &
$0.176_{-0.002}^{+0.003}$ &  & $ 2.54_{-0.22}^{+0.07}$ & $0.500_{-0.000}^{+0.005}$ &   & 2.08 & 1315.83/1164 \\

 & Combined & PN/M1/M2 & 41.0/79.0/80.8 &
$0.159_{-0.004}^{+0.005}$ & & $1.67_{-0.03}^{+0.04}$ & $0.540_{-0.004}^{+0.004}$ &   & 2.18 & 1679.66/1440 \\

\hline
 & & & & &    phabs*(ray$+$diskbb$+$po)       & & & &  \\
\hline

 & Combined & PN/M1/M2 & 41.0/79.0/80.8 &$0.198_{-0.009}^{+0.009}$ &
$0.161_{-0.016}^{+0.015}$ & $1.16_{-0.03}^{+0.04}$ & & $2.30_{-0.04}^{+0.05}$ & 2.45 & 1574.66/1435 \\

\hline
 & & & & &    phabs*(ray$+$diskpbb)      & & & &  \\
\hline

 & Combined & PN/M1/M2 & 41.0/79.0/80.8 &$0.172_{-0.005}^{+0.006}$ &
$0.153_{-0.013}^{+0.020}$ & $1.70_{-0.04}^{+0.03}$ & $0.535_{-0.004}^{+0.004}$ &   & 2.23 & 1578.18/1437 \\

\hline
 & & & & &    phabs*(ray$+$diskpbb$+$po)       & & & &  \\
\hline

 & Combined & PN/M1/M2 & 41.0/79.0/80.8 &$0.166_{-0.014}^{+0.007}$ &
$0.164_{-0.013}^{+0.009}$ & $1.38_{-0.08}^{+0.09}$ & $0.571_{-0.030}^{+0.032}$ & $2.09_{-0.12}^{+0.11}$ & 2.23 & 1539.53/1434 \\

\hline \\
\end{tabular}
\begin{minipage}{180mm} \normalsize
Instruments: data from which instrument, PN, MOS1(M1) or MOS2(M2),
are used; Exposure: clean exposures for corresponding instruments
after background flares excluded; $N_{\rm H}$: column density
along the line of sight; $kT$: plasma temperature; $kT_{\rm in}$:
inner disk temperature; $p$: the temperature gradient; $\Gamma$:
power-law photon index ; $f_{\rm X}$: 0.3--10 keV intrinsic flux
in the units $10^{-11}$ ergs cm$^{-2}$ s$^{-1}$; $\chi^2$/dof:
$\chi^2$ and degree of freedom for the best-fit model. \\ \quad
\quad Combined: data from combining all observations except for
Obs \#4 and \#8.
\end{minipage}
\end{table*}
\end{center}

\end{document}